\documentclass[conference]{IEEEtran}

\usepackage{amssymb,amsmath}%,amsthm}
\usepackage{graphicx,epstopdf,subfigure}
\usepackage{citesort}

\newtheorem{theorem}{Theorem}
\newtheorem{lemma}{Lemma}
\newtheorem{definition}{Definition}
\newtheorem{problem}{Problem}

\newcommand{\mcal}[1]{{\mathcal{#1}}}
\newcommand{\mbb}[1]{{\mathbb{#1}}}
\newcommand{\mbf}[1]{{\mathbf{#1}}}

\newcommand{\field}[1]{\mbb{#1}}

\renewcommand{\r}{{\mathbf{r}}}
\renewcommand{\k}{{\mbf{k}}}
\renewcommand{\j}{{\iota}}

\renewcommand{\S}{\field{S}}
\renewcommand{\L}{\mcal{L}}
\newcommand{\num}{\mcal{D}}
\newcommand{\threshold}{\mcal{N}}

\newcommand{\conj}[1]{{\overline{#1}}}
\newcommand{\abs}[1]{{\left|#1\right|}}

\newcommand{\trunc}{{D}} 	%truncation
\newcommand{\band}{{B}}		%bandlimit
\newcommand{\wave}{{W}}		%wavefield
\newcommand{\helm}{{G}}		%wavefield Helmholtz
\graphicspath{{.}{./figs/}}
\centerfigcaptionstrue

\newcommand{\footnotethanks}[1]{%
\renewcommand{\thefootnote}{\empty}%
\footnotetext{#1}%
\renewcommand{\thefootnote}{\arabic{footnote}}
\setcounter{footnote}{0}}%

\begin{document}

\title{Space-Time-Frequency Degrees of Freedom: \\ Fundamental Limits for Spatial Information}

\author{
	\authorblockN{Leif W. Hanlen$^\dag$ and Thushara D. Abhayapala$^\dag$}%
	\authorblockA{National ICT Australia and Australian National University\\
	{\tt \{Leif.Hanlen,Thushara.Abhayapala\}@nicta.com.au}}
}
 
\maketitle

\begin{abstract}
We bound the number of electromagnetic signals which may be observed over a frequency range $[F-W,F+W]$ a time interval $[0,T]$ within a sphere of radius $R$. We show that the such constrained signals may be represented by a series expansion whose terms are bounded exponentially to zero beyond a threshold. Our result implies there is a finite amount of information which may be extracted from a region of space via electromagnetic radiation.
\end{abstract}

\section{Introduction}

\footnotethanks{${}^\dag$ L. W. Hanlen and T. D. Abhayapala also hold appointments with the Research School of Information Sciences and Engineering, ANU. National ICT Australia is funded through the Australian Government's \emph{Backing Australia's Ability initiative}, in part through the
Australian Research Council.}

Wireless communication is fundamentally limited by the physics of the medium. Electromagnetic  wave propagation has been given~\cite{Hui01,Marzetta2002} as a motivation for developing such limits: information is ultimately carried on electromagnetic waves. Narrowband degrees of freedom (dimensionality)  results have been given for dense multipath~\cite{JonKenAbh:ICASSP02,Kennedy:0203b,PoonBroTse2005} and subsequently extended to sparse systems. Here, the signal bandwidth is negligible: dimensionality results are defined in  wavelengths.

Narrow-band wavefields were shown to have limited  concentrations~\cite{PollAbhaKenn:ISIT2004}. The limit was based upon the free-space Helmholtz (wave) equation -- a time independent variation of the electromagnetic  wave~\cite{ArfkenWeber2005}. Such waves may be represented by a functional series, whose  terms are bounded exponentially toward zero beyond some limit. This limit was used to describe a random MIMO channel in dense~\cite{KennedyAbhaya:2004} multipath and provide capacity results.

More recent work -- including  wide-band MIMO motivates analysis of the capability of spatially diverse signals to support multiplexing over significant bandwidths. 

\begin{quote}
\emph{Given a region, bounded by radius $R$,  centre-frequency $F$, bandwidth $2W$ and observation time $T$, what is the number $\num$ of wireless (electromagnetic) signals may be observed?}
\end{quote}

In~\cite{HanAbh:ausctw2007}  an approximate dimensionality result was given. This bound suffered was excessively complex -- resulting in a loose over-bound. 
In this work we provide exponential error bounds -- reflecting the results of~\cite{JonKenAbh:ICASSP02}, and provide a tighter bound on the dimensionality of 3D-spatial broadband signals.
In developing our result we will also show how the four parameters $(R,T,W,F)$ may be traded against each other. 

The remainder of this paper is arranged as follows: We provide a truncation point and bound the error for electromagnetic signals in space in Section~\ref{S:dimension}. This gives our main result in Theorem~\ref{thm:3d}. Section~\ref{S:plots} gives plots of the degrees of freedom, while Section~\ref{S:example} provides a simple application of the dimensionality result to MIMO mutual information. We draw conclusions in Section~\ref{S:conclusion}.
We formally define spatially constraining signals as a compact operator, and develop proofs in the Appendix.

\section{Dimensionality}\label{S:dimension}

Existing dimensionality results for signals 3D space, with non-trivial bandwidth are limited by
\begin{align}
\num_{WT}&=2WT+1\label{E:Shannon}
\\
\num_{space}&=\left(\left\lceil\frac{e\pi R F}{c}\right\rceil+1\right)^2\label{E:Kennedy3d}
\end{align}
where \eqref{E:Shannon} is from~\cite{Shannon1949}, formalised in~\cite{Pollak0161} and \eqref{E:Kennedy3d} is from~\cite{Kennedy0202}. Fundamentally, we seek to develop a result which combines both \eqref{E:Shannon} and \eqref{E:Kennedy3d}: broadband, spatially diverse signals.
%\section{Model}

Source-free (propagating) electric fields $\Psi(\r,t)$, are solutions of the free-space Maxwell wave equation~\cite{ArfkenWeber2005}:
\begin{equation}\label{E:maxwell}
\left(\triangle - \frac{1}{c^2} \frac{\partial^2}{\partial t^2}\right)\cdot\Psi(\r,t)=0
\end{equation}
where $\triangle=(\partial^2/\partial x^2, \partial^2/\partial y^2, \partial^2/\partial z^2)$ 
and \mbox{$c=3\times10^8\textrm{ms}^{-1}$} is the speed of light. The vector $\r=(r,\theta,\phi)$ with $0\leq\theta<\pi$, $0\leq\phi<2\pi$ denotes position. We now formally pose:% in three dimensions.
\begin{problem}\label{Prob:1}
\begin{itshape}
Given a function in space-time $x(\r,t)$ which is non-zero for $\|\r\|\leq R$ and $t\in[0,T]$  has a frequency component in $[F-W,F+W]$ and satisfies \eqref{E:maxwell}; what number $\mcal{D}$ of signals $\varphi(\r,t)$ are required to parameterize $x(\r,t)$?
\end{itshape}
\end{problem}

%%% Critical Threshold idea
Before addressing Problem~\ref{Prob:1}, observe that we may represent any signal which is constrained to $[0,T]\times[0,R]$ and satisfying \eqref{E:maxwell} by a solution of \eqref{E:maxwell}. In Cartesian coordinates such solutions are exponentials~\cite{Gradshteyn00}:
\begin{equation}\label{E:Psi-def}
\Psi_{\k}(\r,t) = 
\exp\left(-\j |\k|ct - \j \k\cdot \r\right)
\end{equation}
where $\k$ is the vector wave-number and $\j=\sqrt{-1}$. The magnitude $|\k|=2\pi f/c$ is the well-known scalar wave-number.

Any function represented in terms of \eqref{E:Psi-def} may be written as a series expansion~\eqref{E:Psi-series}. The series itself is not important, simply that it exists. This series may be truncated at a point $\threshold$ which is an increasing function of frequency, time and space:
\begin{lemma}[Truncation Point]\label{lem:truncation-point}
\begin{itshape}
The critical threshold (a function of $|\r|$, $t$) at frequency $f$ is
\begin{align}
\threshold(\r,t;f) &= \threshold_T(t;f)+\threshold_\S(\r;f) 
\\
&\threshold_T(t;f)= \left\lceil e \pi \Delta_f t\right\rceil
\label{E:constraint:freq}
\\
&\threshold_\S(\r;f)= \left\lceil e \pi f\frac{ |\r|}{c} \right\rceil
\label{E:constraint:space}
\end{align}
\end{itshape}
\end{lemma}
\smallskip
Once we have chosen an appropriate threshold, increasing $\threshold$ reduces the truncation error exponentially.
\begin{lemma}[Truncation Error]\label{lem:truncation}
\begin{itshape}
The truncation error is 
\begin{equation}
\epsilon_{\threshold}\leq  2/e \approx 0.74
\end{equation}
 and for any $\alpha,\delta\geq1$ 
\begin{equation}
\epsilon_{\threshold+\delta +\alpha} < \epsilon_{\threshold }\cdot e^{2-\delta-\alpha}
\end{equation}
\end{itshape}
\end{lemma}
\smallskip

We may use \eqref{E:constraint:freq} and \eqref{E:constraint:space} and the tightness of the truncation to define the dimesionality of the 3D spatial signals.
%==================================================
\begin{theorem}[3D dimensionality $\mcal{D}_{3D}$]\label{thm:3d}
The number of orthogonal electromagnetic waves which may be observed in a three-dimensional spatial region  bounded by radius $R$, over frequency range $F\pm W$ and time interval $[0,T]$ is 
\begin{multline}\label{E:thm2}
\mcal{D}_{3D}\approx 
 (e\pi2WT+1)\left(\frac{e\pi R (F-W)}{c}+1\right)^2
\\ +
 e\pi2WT\left(\frac{e\pi R }{c}\right)^2\left[2FW-\frac{2}{3}W^2\right]
 \\
 +\left(\frac{e\pi R }{c}\right)^24FW
 +e\pi2WT \left[\left(\frac{e\pi R }{c}\right) F+\frac{1}{6}\right]
\end{multline}
\end{theorem}
\smallskip

Note $F\geq W$ by definition, so both terms in \eqref{E:thm2} are always positive. 
The second term in \eqref{E:thm2} gives an estimate of the effect of signal bandwidth on the degrees of freedom in space. For practical applications we expect $R\ll c$ and  $TW$ to be small ($TW< 10$). Thus, for the second term in \eqref{E:thm2} to be non-negligible we require  $\sqrt{FW} \approx c$, where we are interested only in magnitudes.

\subsection{Asymptotic Results}

For 
$R\to0$ \eqref{E:thm2} reduces to $7e\pi TW/3+1$ which over-bounds \eqref{E:Shannon}.
For
$T,W\to0$,  \eqref{E:thm2} reduces to \eqref{E:Kennedy3d} while for \mbox{$W\neq0, T\to0$} \eqref{E:thm2} reduces to \eqref{E:Kennedy3d} with $F\to F+W$. 
For extreme broadband signals $F=W$, and 
\begin{equation}
\num_{3D} \propto 2e\pi TW \left(\frac{e\pi R }{c}\right)^2 \frac{4}{3}W^2
\end{equation}
and when all parameters are non-trivial
\begin{equation}
\num_{3D} \propto 2e\pi TW \left(\frac{e\pi R }{c}\right)^2 \left(F^2+\frac{W^2}{3}\right)
\end{equation}

\section{Plots}\label{S:plots}

\begin{figure}
\begin{center}
\setlength{\unitlength}{0.0040\textwidth}
\subfigure[DoF vs bandwidth $W$ and region radius $R$\label{F:largeWa}]{%
\begin{picture}(110,80)
\put(0,0){\includegraphics[width=0.40\textwidth]{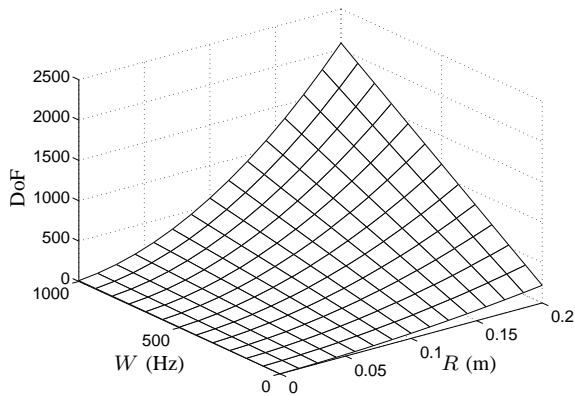}}
\put(75,5){\footnotesize{$R$ (m)}}
\put(15,5){\footnotesize{$W$ (Hz)}}
\put(-4,35){\rotatebox{90}{\footnotesize{DoF}}}
\end{picture}
}
\subfigure[Contours\label{F:largeWb}]{ %
\begin{picture}(110,80)
\put(0,3){\includegraphics[width=0.40\textwidth]{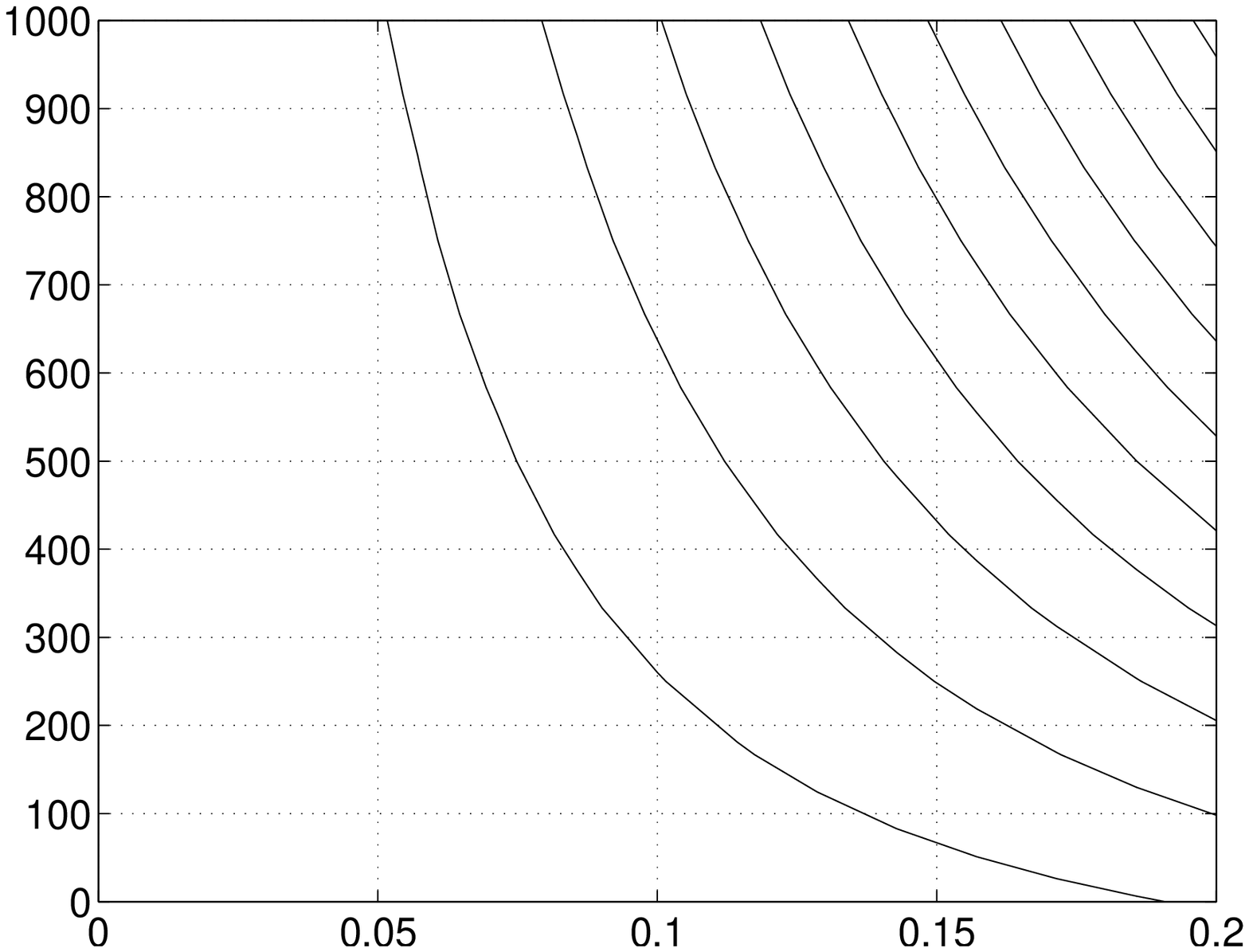}}
\put(45,-2){\footnotesize{$R$ (m)}}
\put(-3,35){\rotatebox{90}{\footnotesize{$W$ (Hz)}}}
\end{picture}
}
\caption{Number of degrees of freedom for moderate $W$ and $T=0.5ms$, $F=2.4GHz$, $\lambda=0.125m$: Curvature toward bottom of Fig.~\ref{F:largeWb} denotes saturation wrt. radius.}\label{F:largeW}
\end{center}
\end{figure}

We have considered two common spatially diverse scenarios. In Figure~\ref{F:largeW} we have used a centre frequency $F=2.4$GHz, 1kHz bandwidth and $R<2\lambda$. Figure~\ref{F:largeWa} shows a mesh of the degrees of freedom, while Figure~\ref{F:largeWb} gives a contour plot. The super-linear growth in DoF can be seen as both $R$ and $W$ increase.

\begin{figure}
\begin{center}
\setlength{\unitlength}{0.0040\textwidth}
\subfigure[DoF vs bandwidth $W$ and region radius $R$\label{F:smallWa}]{%
\begin{picture}(110,80)
\put(0,0){\includegraphics[width=0.40\textwidth]{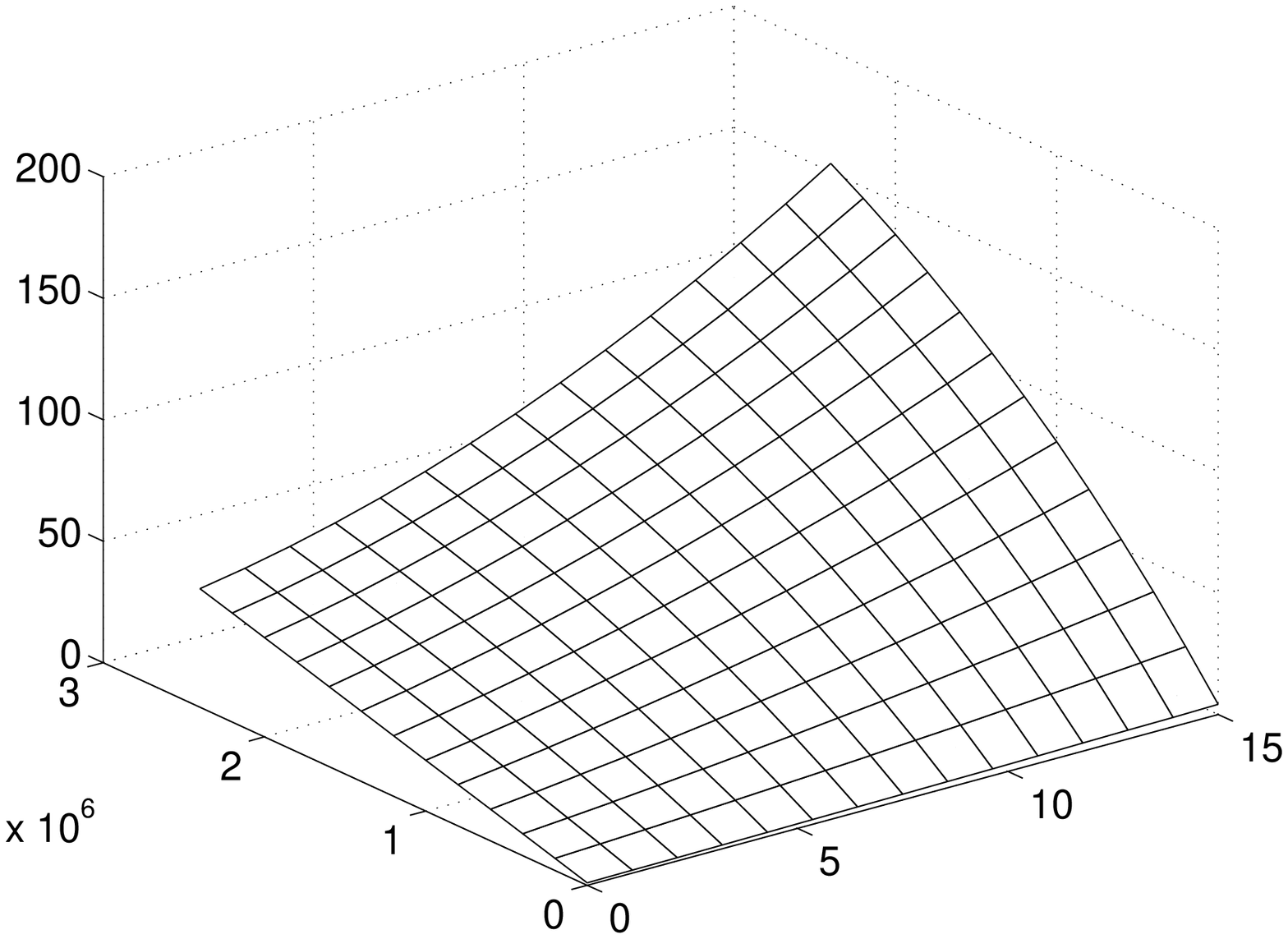}}
\put(75,5){\footnotesize{$R$ (m)}}
\put(15,5){\footnotesize{$W$ (Hz)}}
\put(-2,35){\rotatebox{90}{\footnotesize{DoF}}}
\end{picture}
}
\subfigure[Contours\label{F:smallWb}]{%
\begin{picture}(110,80)
\put(0,3){\includegraphics[width=0.40\textwidth]{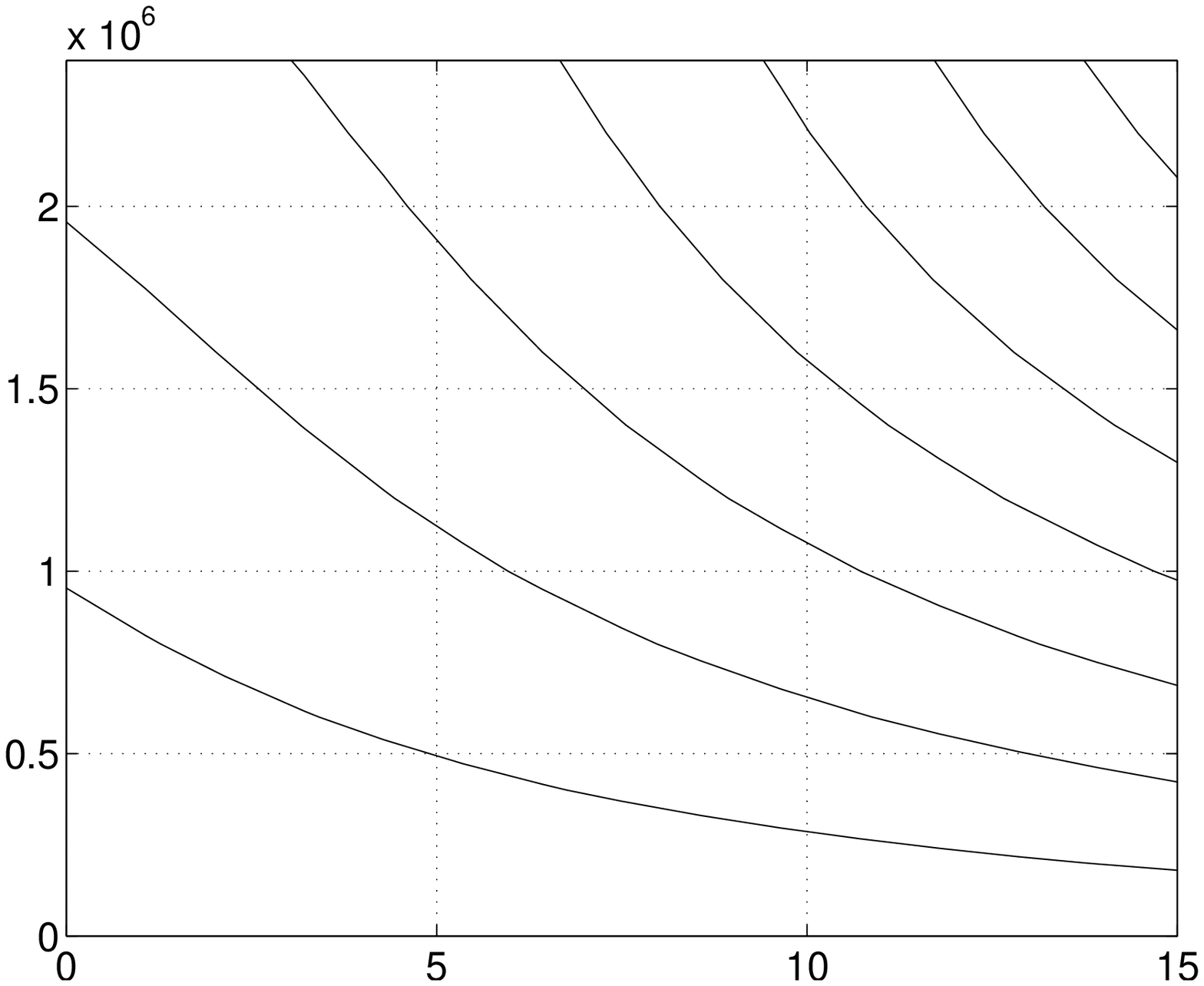}}
\put(45,-2){\footnotesize{$R$ (m)}}
\put(-3,35){\rotatebox{90}{\footnotesize{$W$ (Hz)}}}
\end{picture}
}
\caption{Number of degrees of freedom for large $W$, small $R$ and $T=1\mu s$, $F=2.4MHz$, $\lambda=125m$ }\label{F:smallW}
\end{center}
\end{figure}

Figure~\ref{F:smallW} shows the DoF for a broadband signal $W\leq F$ with centre frequency $F=2.4$MHz. In this case the centre wave-length is 125m (and at $F+W$, $\lambda_{\min}=62.5m$) so $R\ll \lambda/2$. In this case at $R\to0$ we see the usual $2WT$ linear growth in DoF, at $R\approx \lambda_{\min}/4$ we see a knee-point in the DoF surface, corresponding to spatial degrees of freedom becoming effective. This can be seen by the curvature of the contours near $R=5m$ in Figure~\ref{F:smallWb}

\section{Example: Mutual Information $\mcal{I}$}\label{S:example}
We assume all channel eigenvalues are equal magnitude in space and frequency, up to $\threshold_T$ and $\threshold_S$ respectively. The transmitter  uses  \eqref{E:Psi-def} as matched filters for the channel and sends uniform power $\rho$ on the subset of modes with non-neglible magnitude. This is a reasonable capacity approximation~\cite{HanlTimo:ITW2006}.

A na\"ive application of \eqref{E:thm2} to mutual information  would be $\mcal{I}=\num \log(1+\rho/\num) \leq \rho$. This ignores the random nature of the MIMO channel:  spatial signals are mixed through a random scattering channel, while frequency signals are not. Assume both transmitter and receiver have identical geometries and are situated in dense scatter.

Consider Figure~\ref{F:geometric-trap}, every horizontal dashed line  may be considered as an input element with  each element operating at $2W$ different frequency taps (eg. through OFDM). At frequency $F-W\leq f\leq F+W$ there are $N_t=(e\pi R f/c+1)^2$,  independent input signals from~\eqref{E:Kennedy3d}. Then there are 
\begin{equation}\label{E:MIMO-total}
N=\sum_{f} N_t(f) = \sum_{f=F-W}^{F+W}(e\pi R f/c+1)^2
\end{equation}
parallel input channels and mutual information at $f$ is $\mcal{I}_f$
\begin{equation*}
\mcal{I}_f = \log\det\left(I_{N_t} + \frac{\rho}{N}XX^*\right)
%\\
\to N_t \log\left(1+P \frac{N_t}{N}\right)
\end{equation*}
Where $X$ is a Gaussian random matrix of dimension $N_t(f)$, see for example~\cite{PollAbhaKenn:ISIT2004}.
Since each frequency channel is independent, the total mutual information is found by combining \eqref{E:MIMO-total} and
\begin{equation}\label{E:thesum}
\mcal{I} = \sum_{f=F-W}^{F+W} N_t(f)\log\left(1+\rho\frac{ N_t(f)}{N}\right)
\end{equation}
which may be calculated numerically. Note, in the case of $N_t(f)=\mathrm{const}$ we return to the classic parallel channel result $\mcal{I} = \rho N_t$. For $N_t(f)$ an increasing function of $f$, the sum \eqref{E:thesum} is increased, thus 
\begin{equation}
\mcal{I}\geq \rho \left(\frac{e\pi R (F-W)}{c}+1\right)^2
\end{equation}
Due to random scattering, spatial modes provide a linear increase in capacity, while frequency modes provide parallel channels.

\section{Conclusion}\label{S:conclusion}

We have shown that the degrees of freedom for a spherically restricted broadband wireless signal is proportional to the surface area of the spatial region, the square of the frequency  and bandwidth and the DoF of the broadband signal itself. We have shown that the error associated with truncating such a signal at $N$ terms decreases exponentially as $N$ increases.

As an example we have shown that broadband spatial communication systems may have a capacity beyond that expected by combining MIMO capacity results by with parallel frequency channels.

%=================================================================
%=================================================================
\appendices

\section{Operator Material}\label{S:operator}

\begin{definition}[Truncation Projection $\trunc$]\label{D:trunc-def}
\begin{itshape}
Given the spatial interval $\S$ and time interval $[0,T]$, the truncation operator $\trunc$ sets a field $f(\r,t)$ to zero outside the time- and space- intervals.
\begin{equation}\label{E:trunc-def}
%f'(\r,t)=
\trunc f(\r,t) = \begin{cases}
f(\r,t) & \r\in\S \text{ and } t\in [0,T]
\\
0 &\text{else}
\end{cases}
\end{equation} 
\end{itshape}
\end{definition}
\bigskip

\begin{definition}[Wavefield Projection $\wave$]\label{D:wave-def}
\begin{itshape}
The wavefield projection $\wave$ projects a field $f(\r,t)$ onto solutions $\Psi(\r,t)$ of~\eqref{E:maxwell}.
\begin{align}
\wave f(\r,t) \label{E:wave-def}
 &= \sum_{l,m,n} f_{l,m,n} \Psi_{l,m,n}(\r,t) 
\\
&f_{l,m,n} = \iint f(\r,t) \conj{\Psi_{l,m,n}(\r,t)} \, dt \, d\r \notag
\end{align} 
$\k$ is the wave vector in three dimensions $\k=(k_x,k_y,k_z)$ with scalar wave number $k=|\k|=(k_x^2+k_y^2+k_z^2)^{1/2}$ \cite[eqn.6.94 p.759]{Gradshteyn00}. $\Psi_{l,m,n}$ is given by \eqref{E:Psi-def} with the values of $k$ chosen discretely:
\begin{equation}
k = \frac{2\pi}{cT}p \quad k_x = \frac{2\pi}{L}l \quad k_y = \frac{2\pi}{L}m \quad k_z = \frac{2\pi}{L}m 
\end{equation}
\end{itshape}
\end{definition}
\bigskip

Using \eqref{E:trunc-def} and \eqref{E:wave-def} we may write the signal observed in $\S\times T$ as:
\begin{equation}\label{E:WD-operator}
g(\r,t) = \wave \trunc f(\r,t)
\end{equation}

\begin{lemma}\label{L:compact}
$\wave \trunc$ is a compact operator.
\end{lemma}
Lemma~\ref{L:compact} emphasizes that although there are infinitely many independent electromagnetic waves, only a finite number of electromagnetic signals may be resolved within the region \mbox{$\S\times T$}. The implication of this is that any approximation for a given signal has a bounded error.

\section{Proofs}

\begin{proof}[Lemma~\ref{lem:truncation-point}]
%=================================
%Truncation

From~\eqref{E:Psi-def}
\begin{align}
\Psi_\k(\r,t) &=\exp(-\j k_{\min}ct)\exp\left(-\j \acute{k}ct - \j \k\cdot \r\right)
 \\
&=\exp(-\j k_{\min}ct)\hat{\Psi}_\k(\r,t)
\end{align}
where $\k\cdot\r$ denotes the vector dot product and \mbox{$0\leq \acute{k}\leq 4\pi W/c$}. We wish to bound the number of terms required to approximate this function. Note $\exp(-\j k_{\min}ct)$ has exactly one degree of freedom, so we may equivalently calculate the DoF for $\hat{\Psi}$.
 %
%-- first half
Using the Jacobi-Anger expansion \cite[eqn2.45, pp.32]{Colton98} and summation theorem \cite[eqn2.29, pp.27]{Colton98}
\begin{equation}\label{E:jacobi}
e^{-\j \k\cdot \r} = 4\pi\sum_{n=0}^\infty \j^n j_n(k\|\r\|)\sum_{m=-n}^nY_n^m(\hat{\r})\overline{Y_n^m(\hat{\k})}
\end{equation}
where $j_n(z)=\sqrt{\frac{\pi}{2z}}J_{n+\frac{1}{2}}(z)$ is a spherical Bessel function. 

From~\cite[8.534.1]{Gradshteyn00}
\begin{equation}\label{E:sum-rule}
e^{-\j c \acute{k} t}=\sum_{p=0}^\infty \j^p(2p+1) j_p(c\acute{k}t)
\end{equation}
Combining \eqref{E:jacobi} and \eqref{E:sum-rule}
\begin{multline}\label{E:Psi-series}
\Psi^{(P,N)}(\r,t)=4\pi\sum_p^P\j^p(2p+1)j_p(c\acute{k}t)\\ \times \sum_{n,m}^N\j^nj_n(k\|\r\|)Y_n^m(\hat{\r})\overline{Y_n^m(\hat{\k})}
\end{multline}

We may truncate $\Psi(\r,t)$ and bound the error as in~\cite{JonKenAbh:ICASSP02}:
\begin{align*}
\epsilon_{(P,N)} &= \left|\Psi(\r,t)-\Psi^{(P,N)}(\r,t)\right|
\\
&=4\pi\left|\sum_{p>N}^\infty\j^p(2p+1)j_p(c\acute{k}t)\right.
\\
&\ \qquad\ \qquad\times\left.\sum_{n>N}^\infty\j^nj_n(k\|r\|)\sum_{m=-n}^{n}Y_n^m(\hat{\r})\overline{Y_n^m(\hat{\k})}\right|
\end{align*}
Taking the absolute values inside the summation and using $\left|\sum_{m=-n}^nY_n^m(\hat{\r})\overline{Y_n^m(\hat{\k})}\right| \leq (2n+1)/(2\pi)$ \cite[pp.27]{Colton98} gives
\begin{equation}
\epsilon_{P,N}\leq 2 \sum_{p>N}^\infty(2p+1)\abs{j_p(c\acute{k}t)}\sum_{n>N}^\infty(2n+1)\abs{j_n(k|r|)}
\end{equation}

From~\cite{Kennedy0202,JonKenAbh:ICASSP02}
\begin{equation}
\abs{j_n(x)}\leq \frac{\sqrt{\pi}}{2}\frac{1}{\Gamma(n+3/2)}\left(\frac{x}{2}\right)^n 
\end{equation}
Using the identity \mbox{$(2n+1)/\Gamma(n+3/2)=1/\Gamma(n+1/2)$} and~\cite[pp.257]{Abramowitz72} \mbox{$\Gamma(n+1/2)>e^{-n-1/2}(n+1/2)^{n}(2\pi)^{1/2}$} 
\begin{align}
\epsilon_{P,N}&<\frac{e}{2}\sum_{p>P}\left[\frac{(c\acute{k}t)e}{2(p+1/2)}\right]^p\sum_{n>N}\left[\frac{(k|\r|)e}{2(n+1/2)}\right]^n
\label{E:series}
\\
&<\frac{e}{2}\sum_{p>P}\left[\frac{(c\acute{k}t)e}{2(P+1)}\right]^p\sum_{n>N}\left[\frac{(k|\r|)e}{2(N+1)}\right]^n\label{E:trunc-bound}
\end{align}
Both sums \eqref{E:trunc-bound} converge if \mbox{$P+1>ec\acute{k}t/2$} and \mbox{$N+1>ek|\r|/2$}.
\begin{align}
\epsilon_{P,N}&<\frac{e (N+1)(P+1)%
 \left(\frac{ec\acute{k}t}{P+1}\right)^P
 \left(\frac{ek|\r|/c}{N+1}\right)^N
 }{(2N+2-x)(2P+2-y)2^{N+P-1}}
 \\
 &<2e
  \left(\frac{e\pi \acute{f} t}{P+1}\right)^P
 \left(\frac{e\pi f |\r|/c}{N+1}\right)^N\label{E:product-bound}
\end{align}
$0\leq\acute{f}\leq2W$ and $F-W\leq f\leq F+W$
\end{proof}

\begin{proof}[Lemma~\ref{lem:truncation}]
Define $x=eckt$ and $y=ekR$ then using~\eqref{E:product-bound} 
\begin{multline*}
\frac{\epsilon_{N+\alpha,P+\delta}}{\epsilon_{N,P}}%
=\left(\frac{P+1}{P+1+\delta}\right)^{P+\delta}\left(\frac{N+1}{N+1+\delta}\right)^{N+\alpha}
\\
\times \left(\frac{x}{P+1}\right)^\delta \left(\frac{ y }{N+1}\right)^\alpha
\end{multline*}
Now $\alpha,\delta\geq1$, $P+1>x$ and $N+1>y$ by definition
\begin{equation*}
\frac{\epsilon_{N+\alpha,P+\delta}}{\epsilon_{N,P}}%
\leq \left(\frac{P+1}{P+\delta}\right)^{P+\delta}\left(\frac{N+1}{N+\delta}\right)^{N+\alpha}
\leq e^{1-\delta}e^{1-\alpha}
\end{equation*}
Using the identity $(1+a/N)^N<e^{a}$
\end{proof}

\begin{proof}[Theorem~\ref{thm:3d}]
From~\eqref{E:product-bound} $P$ may found directly by using the maximum values of $t$ and $\acute{k}$:
\begin{equation}
P=e\pi2 W T
\end{equation}
Truncating \eqref{E:product-bound} at $P$ gives $\num_T=2e\pi W T+1$ terms. The same technique cannot be used for $N$, since the bound becomes excessively loose. Instead we use a geometric argument:

\begin{figure}
\centering 
\setlength{\unitlength}{0.50mm}
\begin{picture}(100,90)
\put(0,3){\includegraphics[width=50mm]{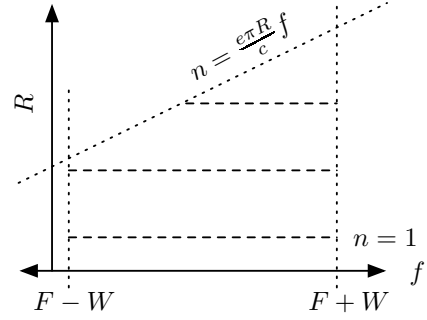}}
\put(5,-2){$F-W$}
\put(78,-2){$F+W$}
\put(105,6){$f$}
\put(0,50){\rotatebox{90}{$R$}}
\put(45,59){\rotatebox{28}{$n=\frac{e\pi R}{c}f$}}
\put(90,15){$n=1$}
\end{picture}
\caption{Geometry for spatial functions. $F-W\leq f\leq F+W$ and $0\leq |\r|\leq R$. Time forms a third dimension (into the page).}\label{F:geometric-trap}
\end{figure}

Consider Figure~\ref{F:geometric-trap}. The dotted vertical lines give the frequency constraints, while the top line $n=e\pi R f/c$ gives the spatial constraint of \eqref{E:constraint:space}. The constraint set defines a trapezium in the space-frequency plane (or trapezoid when time is included). The heights of the trapezium are found from \eqref{E:constraint:space} with $f=F\pm W$. Note that the figure is not drawn to scale: for any reasonable value of $R$, the slope of the line $e\pi R/c$ is almost zero. 

Each horizontal dashed line represents a collection of time-frequency signals which may be observed at a point in space. Spatial diversity allows observation of multiple collections. Each collection has an intrinsic dimensionality of $e\pi 2\Delta_W T$ where $\Delta_W$ is the effective (frequency) bandwidth of the spatial observation. The dimensionality result is obtained by counting each collection, with appropriate dimensionality for each.

The collections are enumerated by the spatial degrees of freedom $n$. 
Each collection is scaled by the spherical Bessel function $j_n(k|\r|)$ from \eqref{E:jacobi}. These functions have a natural high-pass characteristic: for  $k|\r|<n$ then $j_n(k|\r|) \to 0$.
At low values of $n$, 
\begin{equation}\label{E:N0}
0\leq n\leq e\pi R (F-W)/c=N_0
\end{equation}
 each Bessel function is already active and thus each collection has the full $e\pi2\Delta_WT=e\pi2WT$ degrees of freedom. For higher values of $n$, 
\begin{equation}\label{E:N1}
 N_0=e\pi R (F-W)/c<n\leq e\pi R (F+W)/c=N_1
\end{equation}
  each collection has a reduced DoF, since the Bessel functions are only activated part-way through the frequency band.
\begin{equation*}
N_{TW(n)}=
\begin{cases}
2e\pi WT+1 &0\leq n<N_0
\\
e\pi\left(F+W-\frac{c}{e\pi R}n\right)^{+}T+1 &N_0\leq n\leq N_1
\end{cases}
\end{equation*}

For each value of $n$ in Figure~\ref{F:geometric-trap}, there are $2n+1$ independent spatial modes. 
Thus the total degrees of freedom is given by
\begin{align}\label{E:sum-modes}
\num &= \sum_{n=0}^{N_1} N_{TW(n)} (2n+1) \leq \num_{1}+\num_2
\\
\num_1&=(2e\pi WT+1)\sum_{n=0}^{N_0}(2n+1)\label{E:D1}
\\
\num_2&=\sum_{n=N_0}^{N_1}\left[e\pi(F+W)T-\frac{cT}{R}n+1\right](2n+1)\label{E:D2}
\end{align}

We may solve \eqref{E:D1}:
\begin{equation}\label{E:D1soln}
\num_1=\left(2e\pi WT+1\right)\left(\frac{e\pi R}{c}(F-W)+1\right)^2,
\end{equation}
Evaluating \eqref{E:D2} and combining with \eqref{E:D1soln} gives the result.
\end{proof}

%========================================================
\begin{proof}[Lemma~\ref{L:compact}]
We are considering bandlimited electromagnetic signals, in this case $\wave$  may be decomposed into a band-limiting projection $\band$~\cite{Pollak0161} and a spatial wavefield (or Helmholtz) projection $\helm$~\cite{KennedyAbhaya:2004}.
\begin{equation}
\band f(\r,t) = \frac{1}{2\pi}\int_{F-W}^{F+W}d\omega\, e^{\j\omega t}\int_{-\infty}^\infty dt\,e^{-\j \omega t}f(\r,t)
\end{equation}

\begin{equation}
\helm f(\r,t) = \sum_\k f_\k \exp\left( \j \k\cdot \r\right)
\end{equation}
Write \eqref{E:WD-operator} as:
\begin{equation}
g(\r,t) = \helm \band \trunc f(\r,t)
\end{equation}
We know~\cite{Pollak0161} the operator $\band \trunc$ is compact, since it maps a unit ball in $\L_2$ (finite energy signals) to an essentially finite dimensional ball (of approximate dimension $2WT+1$). From~\cite{Kennedy:0203b} $\helm$ is a projection and thus bounded so \cite[Lem.8.3-2 p.422]{Kreysig78} the product $\helm \cdot \band \trunc$ is compact.
\end{proof}

%\bibliographystyle{IEEEtran}
%\bibliography{IEEEabrv,IEEE_leif_abrv,all_refs_bibdesk}

\end{document}